# On the Accuracy of Atomic-Resolution Electrostatic Measurements in 2D Materials

Rafael V. Ferreira[a,b,c,d,*], Sebastian Calderon V.[e], Paulo J. Ferreira[a,b,*]

[a] INL – International Iberian Nanotechnology Laboratory, Av. Mestre José Veiga s/n, 4715-330 Braga, Portugal

[b] Mechanical Engineering Department and IDMEC, Instituto Superior Técnico, University of Lisbon, Av. Rovisco Pais, 1049-001 Lisboa, Portugal

[c] Departamento de Física de Materiales, Universidad Complutense de Madrid, Pl. de las Ciencias 1, 28040 Madrid, Spain

[d] Instituto Pluridisciplinar, Universidad Complutense de Madrid, P.º de Juan XXIII 1, 28040 Madrid, Spain

[e] Department of Materials Science and Engineering, Carnegie Mellon University, Pittsburgh, Pennsylvania 15213, USA

[*]Corresponding authors: rafael.v.ferreira@ucm.es, paulo.ferreira@inl.int

## Abstract

The use of differential phase contrast (DPC) in scanning transmission electron microscopy (STEM) has shown much promise for directly investigating the functional properties of a material system, leveraging the natural coupling between the electron probe and atomic-scale electric fields to map the electrostatic configuration within a sample. However, the high sensitivity of these measurements makes them particularly vulnerable to variations in both sample properties and the configuration of the instrument, stressing the need for robust methodologies to ensure more accurate analyses. In this work, the influence of key instrumental parameters – probe convergence angle, defocus and two-fold astigmatism – on atomic-resolution segmented-detector DPC-STEM measurements is evaluated through extensive image simulations. Results show that the limit of interpretability for a 21 mrad defocused probe is found at a magnitude of 4 nm, where electrostatic field magnitude can be underestimated by about 16 % in overfocus and just above 10 % in underfocus. Equivalent results for a 30 mrad probe demonstrate underestimated values around 30 % at overfocus and 20 % for underfocus, at a lower interpretability limit of 3 nm. Two-fold astigmatism introduces orientation-dependent variations that surpass 40 % for magnitudes below 3 nm, but a reduction in sensitivity to the aberration is observed when oriented along detector-segment edges. Overall, the analysis confirms the sensitivity and usefulness of the scattergram-based method and underscores the importance of optimized instrumental alignment for accurate CoM-based STEM imaging.

# 1. Introduction

The continual downscaling of electronic devices leads to an increasing susceptibility to defects in the atomic structure of their components, thus enhancing the level of advanced characterization demanded for many applications. As electronic devices advance towards scales commensurate with single point defects, the presence of one or more of these imperfections may have a significant impact on how a material performs during operation. However, the characterization of materials and their defects poses a considerable challenge due to the required spatial resolution combined with the necessity to measure functional properties at very localized scales.

Recently, differential phase contrast (DPC) imaging in a scanning transmission electron microscope (STEM) [1–5] provides a great opportunity to directly investigate defects and their influence at atomic scales. Such observations are based on center-of-mass (CoM) measurements that analyze the overall deflection experienced by the electron beam resulting from its interaction with the intrinsic electromagnetic fields within the sample. In practice, this can be achieved with either detectors split into segments or with the later-developed pixelated detectors (here referred to as DPC- and CoM-STEM, respectively) [5–7], which allow for better resolution in diffraction space.

Because the deflection of the electron beam measured at each scanning point is proportional to the local configuration of electromagnetic forces acting on it, the shift in CoM can be related to the atomic forces sampled by the probe through the material, allowing for the mapping of projected electric fields (cross-correlated with the probe intensity profile) at atomic scales [8]. Following Maxwell's equations, the electrostatic vector field maps (eDPC/eCoM) can then be integrated to retrieve the projected electrostatic potential (iDPC/iCoM) [9–11], which is theoretically proportional to the phase of the electron beam wave function. The integration has the added benefit of suppressing non-conservative field components, so that high-frequency stochastic noise is suppressed in the electrostatic potential images and their resolution is consequently improved. This method has been successfully employed in investigating atomic-scale electrostatic configurations of various materials [12–14], as well as that of dopants and defects [15–18], in addition to examining the dependence of charge distribution on the stacking type of heterostructures [19].

Regarding the experimental setup for CoM-based STEM observations, segmented detectors provide approximated CoM measurements by integrating the intensity captured within large azimuthal ranges and from there estimating its shift from the optical axis. Yet, the accuracy of these detectors is intrinsically limited by their geometry, resulting in a diminished capacity for quantification and a larger uncertainty regarding the electric field direction [7]. In comparison, pixelated detectors capture the intensity distribution of the beam in much greater detail, offering the most accurate CoM measurements currently achievable. Nevertheless, segmented detectors present certain advantages derived from their decreased resolution, which invariably translates into a more manageable amount of data to process. Consequently, while recent technological advancements have pushed pixelated detector acquisition rates to the range of 10 µs/pixel [20,21], segmented detectors are capable of sub-µs acquisitions that retain acceptable signal-to-noise ratios (SNR). For materials that are particularly sensitive to the electron beam, the higher speeds facilitate acquisitions with larger fields of view and add flexibility to capture multiple images in a specific region of the sample before beam damage begins to occur. Furthermore, the lighter processing load allows for real-time DPC-STEM imaging, making it easier to identify *in situ* regions with interesting electromagnetic features, and to better correct electron lens aberrations using the higher sensitivity of the technique [4,22]. These advantages enhance the relevance of segmented detectors in the context of investigating beam-sensitive 2D structures, as demonstrated by the results achieved in several published studies [15–19]. Finally, from a practical standpoint, most modern instruments are now equipped with segmented detectors while pixelated detectors still require significant investments.

Given the high precision required from measurements of atomic electrostatic configurations, the experimental factors involved in CoM-based imaging is an important topic that has been previously explored, namely, the influence of sample and instrument parameters on the features of electrostatic images [23–26]. Among these effects, sensitivity to lens aberrations is particularly significant. In particular, residual two-fold astigmatism, coma, and spherical aberration have been shown to cause distortions and delocalization of imaged atomic cores, while uncorrected defocus can even lead to a complete inversion of the electric field direction. [24,25].

In addition, the thickness of the sample is another aspect with a large impact on CoM-based imaging, as the electron beam continuously broadens while propagating through the material until the linear relationship to the projected field is no longer valid due to the interference from diffraction effects [5,27]. Investigations on this issue have determined a sample thickness limit below 5 nm for avoiding the breakdown of linearity [23], while qualitatively consistent results were shown to be possible for thicknesses up to 15 nm [25]. For this reason, 2D materials present ideal study cases for CoM-based techniques, as the reduced dimensionality induces negligible scattering events and allows for the acquisition of directly interpretable electrostatic maps that are in fair agreement with theoretically predicted values.

In this context, the susceptibility of CoM-based techniques to experimental parameters requires that the influence of each parameter is properly characterized to correctly interpret the experimental observations. Although the use of pixelated detectors allows for ptychographic reconstruction of the electron probe, providing an accurate contrast transfer function with which to offset the influence of aberrations [28,29], the experimental and computational demands of ptychography are still far from a routine method. In addition, considering the benefits and higher availability of segmented detectors, it is relevant to understand the effects of the various parameters on DPC-STEM measurements.

The purpose of this study is to explore how three key experimental parameters in STEM observations – convergence semi-angle ($\alpha$), defocus ($C_1$), and two-fold astigmatism ($A_1$) – affect the retrieval of electric field and potential configurations obtained from CoM-based images. Here, the notation used to identify electron lens aberrations is adopted from Haider et al. [30]. Evaluation of the influence from each instrumental parameter is based on both qualitative changes – how the electrostatic spatial configurations are modified – and quantitative changes – related to the magnitude of electrostatic fields and potentials. This analysis is based on multislice simulations of DPC- and CoM-STEM images, in which the instrumental parameters under study were set to various configurations and evaluated independently. Considering the limitations of CoM-based imaging related to sample thickness, a structural model of monolayer $MoS_2$ was chosen as the object of study, providing an ideal material specimen with a set of characteristics that allows more general conclusions to be drawn, namely a multi-element chemical composition and a layered structure composed by three atomic planes. To assess the influence of the STEM parameters, a baseline is established, free from any lens aberrations, allowing us to determine the expected percentual difference between aberration-induced electrostatic configurations relative to ideal imaging conditions.

This study will help in identifying image features specifically associated with the experimental parameters selected and estimate the corresponding induced error. In this way, a general practical methodology is proposed to analyze experimental results containing effects related to instrumental parameters, without the need for probe-retrieval methods or non-exact deconvolutions. In fact, the implementation of this analysis shows a substantial change in measured electrostatic field configurations due to the addition of defocus, with an underfocused probe affecting the measured spatial distributions, while overfocus is more quantitatively detrimental. In addition, $A_1$ is shown to have a higher impact on the measured configurations, with a modulation induced by its orientation relative to that of the segmented detector. Finally, the integration of the field configurations is shown to have a suppressing effect on the changes induced by the aberrations, as well as on the features associated with the segmented detector geometry.

## 2. Methodology

### 2.1. Computer simulations and instrumental parameters

Multislice image simulations were performed with the Dr. Probe software [31], setting the imaging conditions to match those of a typical STEM instrument, excluding the existence of any lens aberrations apart from those being analyzed. The accelerating voltage was set to 80 kV, considering the damage threshold of $MoS_2$ [32], and an effective source size of 70 pm at FWHM was defined to introduce spatial incoherence effects. The remaining illumination parameters were varied over several simulations in which different configurations were considered. The convergence semi-angle $\alpha$ is analyzed for two typical values for CEOS DCOR correctors – 21 and 30 mrad – with its influence characterized in tandem with that of $C_1$, which is considered within a range of 5 nm to either side of the focal plane, set at the midplane of the $MoS_2$ structure. The non-rotationally symmetric aberration $A_1$ is analyzed up to a magnitude of 3 nm under all possible orientations, and for $\alpha = 30$ mrad, as this configuration offers the best spatial resolution.

### 2.2. Detectors

For the detector geometry used in the simulations, a typical annular four-segment detector installed in an FEI Titan Themis G3 is considered. In addition, while considering that the detector rotation relative to the sample structure is also a factor that will affect the measurements, a standard detector orientation is here defined such that the segments are aligned with the Cartesian axes (Figure S1) to simplify the processing of the DPC and CoM images, aside from when indicated otherwise. The collection angles are defined according to the convergence angle. To maintain similar experimental conditions, the CoM of the CBED patterns captured at each scan point of an image is determined by integrating the intensity distribution up to the maximum collection angle of the corresponding segmented detector configuration. The collection angles, as well as other parameters set for the simulations are summarized in Table 1.

**Table 1.** Values and ranges for parameters used in the simulations of DPC- and CoM-STEM images.

| | | |
|---|---|---|
| **Accelerating voltage** | 80 kV | |
| **Effective source size (FWHM)** | 70 pm | |
| **Convergence semi-angle ($\alpha$)** | 21 mrad | 30 mrad |
| **DPC collection angles** | 3.71–21 mrad | 5.3–30 mrad |
| **Defocus ($C_1$)** | [- 5, 5] nm | |
| **Two-fold astigmatism ($A_1$)** | [0, 3] nm ; [0, 360]° | |

### 2.3. Atomic structural models

The structural model of pristine monolayer $MoS_2$ used in this work is shown in Figure 1. To introduce the effects of thermal diffuse scattering, the image calculations were carried out using frozen lattice approximations, in which the atomic positions in the structure were randomly displaced within 2 % of the interatomic distance. Although the simulated images are based on an isolated atomic potential model, the comparative nature of the analysis means that only differences between images are of concern, making a more realistic atomic potential unnecessary to draw conclusions regarding the influence of experimental parameters.

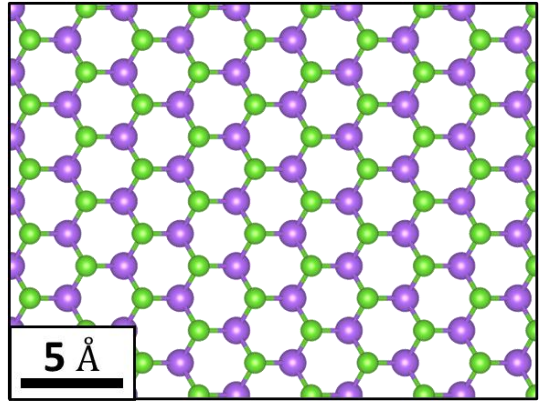

**Figure 1.** Structural model of MoS$_2$ used as input for the multislice simulations in this work. The molybdenum (Mo) and the sulfur (S) atoms are colored in purple and green, respectively.

*2.4. Processing of CoM-based images*

The horizontal and vertical components from simulated DPC- and CoM-STEM images are converted to electrostatic vectorial field images and electrostatic potential images using Python-based self-developed scripts. The electrostatic field is obtained following the linear relation to the beam CoM shift, according to Ehrenfest's theorem [8], while the potential is calculated following Maxwell's laws and the integration scheme discussed in [11].

*2.5. Analysis of electrostatic field and potential images*

As mentioned above, CoM-based images can be processed to retrieve different types of electrostatic quantities, specifically in the form of eDPC/eCoM and iDPC/iCoM images. The vectorial eDPC/eCoM images exhibit complications to direct interpretation and comparison, as they contain both magnitude and direction, and retain the experimental noise that is part of any real observation to varying degrees depending on the sample conditions. On the other hand, the scalar iDPC/iCoM images can be compared directly at the pixel level, facilitating a detailed analysis of the images. In addition, these iDPC/iCoM electrostatic potential images exhibit an improvement in image quality due to noise suppression. In this context, to better understand the impact of the microscopy parameters on the accuracy of atomic-resolution DPC-STEM measurements, it is important to analyze both electrostatic field and potential images.

*2.5.1. Scattergrams*

J. Burger et al. [24] introduced the representation of the vector field in scattergrams – polar graphs where the CoM vectors associated with the pixels of the original eDPC/eCoM image are plotted as points. An example of the two representations is shown in Figure 2 for the case of a MoS$_2$ image simulated with a 30 mrad probe at 80 keV. Figure 2A shows the electrostatic field map, while in Figure 2B, each vector shown in Fig. 2A is represented by a point, with the magnitude of the electrostatic field identically color-coded in both images. In this fashion, the magnitude of the deflections experienced by the electron beam in each pixel is effectively plotted in the scattergrams as distance from the center, whereas their direction is reflected in the azimuthal angle. This representation allows us to readily evaluate the strength of the measured deflection/field in every direction.

Figures 2A and 2B show that the periphery of the scattergram corresponds to the shape of the two types of atomic sites (as indicated by the matching dashed triangles in the electrostatic map and scattergram), with the differing atomic masses translating the crystal symmetry to the scattergram. The core part is then related to the low magnitude regions, namely the interatomic regions and the center of the atomic sites (where electrons are strongly deflected, forming a point of singularity). However, to better determine the influence from any of various experimental parameters, including sample structure and instrumental conditions, a color scale that corresponds to the number of vectors with similar magnitude and direction can be adopted (Figure 2C). This representation shows that the sharply directed transitions from the atomic site regions to the low-magnitude center of the $MoS_2$ hexagonal rings creates points of strong convergence where the vector concentration is highest, resulting in a hexagonal motif at the center of the scattergram.

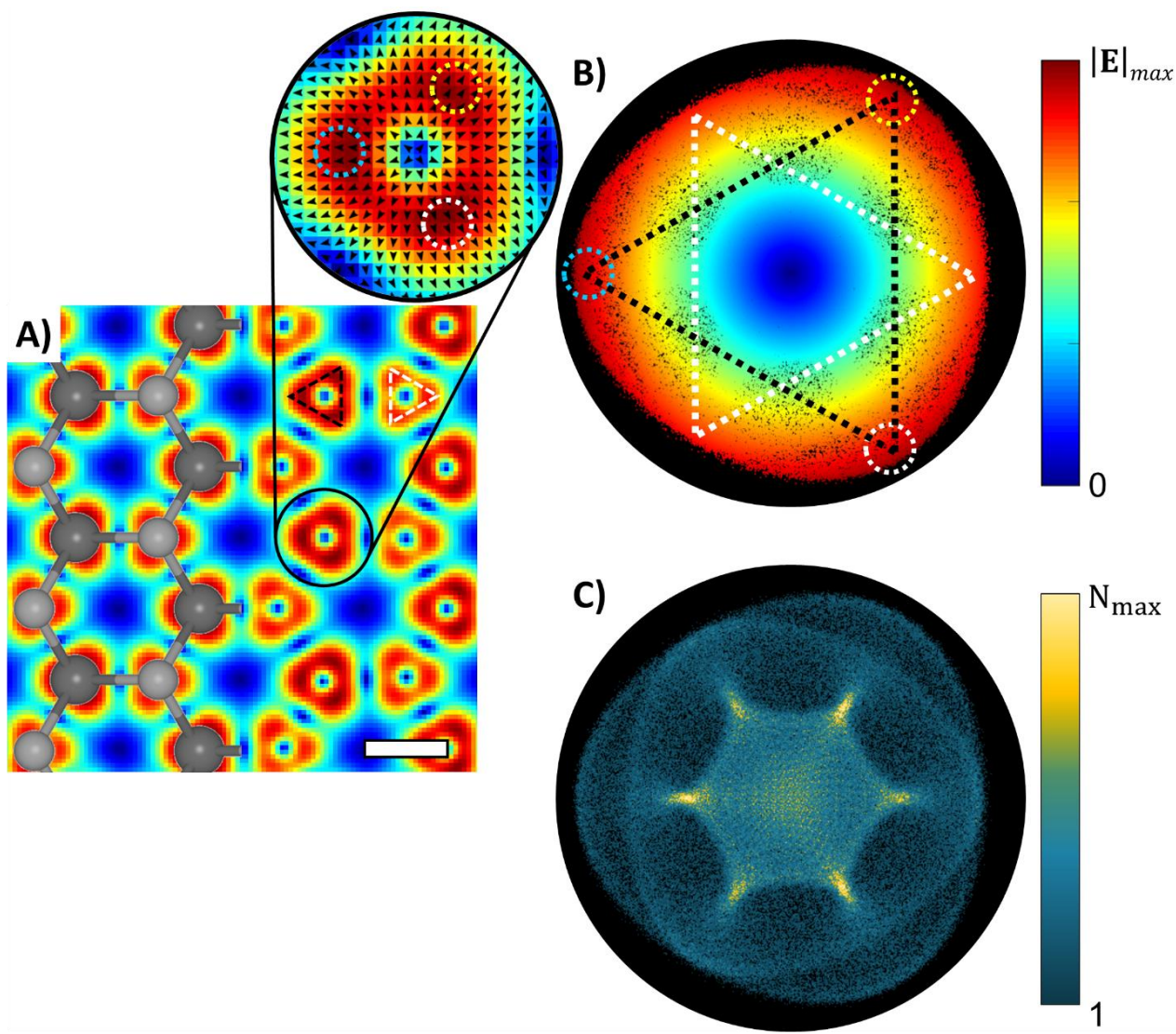

**Figure 2.** A) Structural model of monolayer $MoS_2$ overlaid on the simulated CoM-STEM projected electrostatic field magnitude map, considering an accelerating voltage of 80 kV and a 30 mrad convergence semi-angle. Scale bar corresponds to 2 Å. B) Corresponding scattergram where the electrostatic field magnitude follows the same color scale in both representations. In the magnified region of A), which shows the CoM vectors in each pixel of the image, matching circles indicate where encircled vectors appear as points in the scattergram. C) Monolayer $MoS_2$ scattergram using a concentration-based color scale, indicating the amount of equivalent CoM vectors in A).

To conduct a quantitative analysis and comparison of the information contained in the scattergrams, a slicing-based method is implemented which consists of dividing the full angular range into slices. Following this scheme, scattergrams are sectioned into 24 slices as illustrated in Figure 3, providing sufficient resolution in the azimuthal axis to independently evaluate regions where the intensity remains radially uniform. This slice configuration is also detailed enough to suit the analysis of any other possible projected rotational symmetries. The intensity profiles of each slice plot the magnitude distribution of vectors along its direction, as shown in Figure 3. This distribution can be analyzed for the highest electric field magnitude reached, intensity peaks (points of highest vector concentration), and compared to that of slices in the same or other scattergrams.

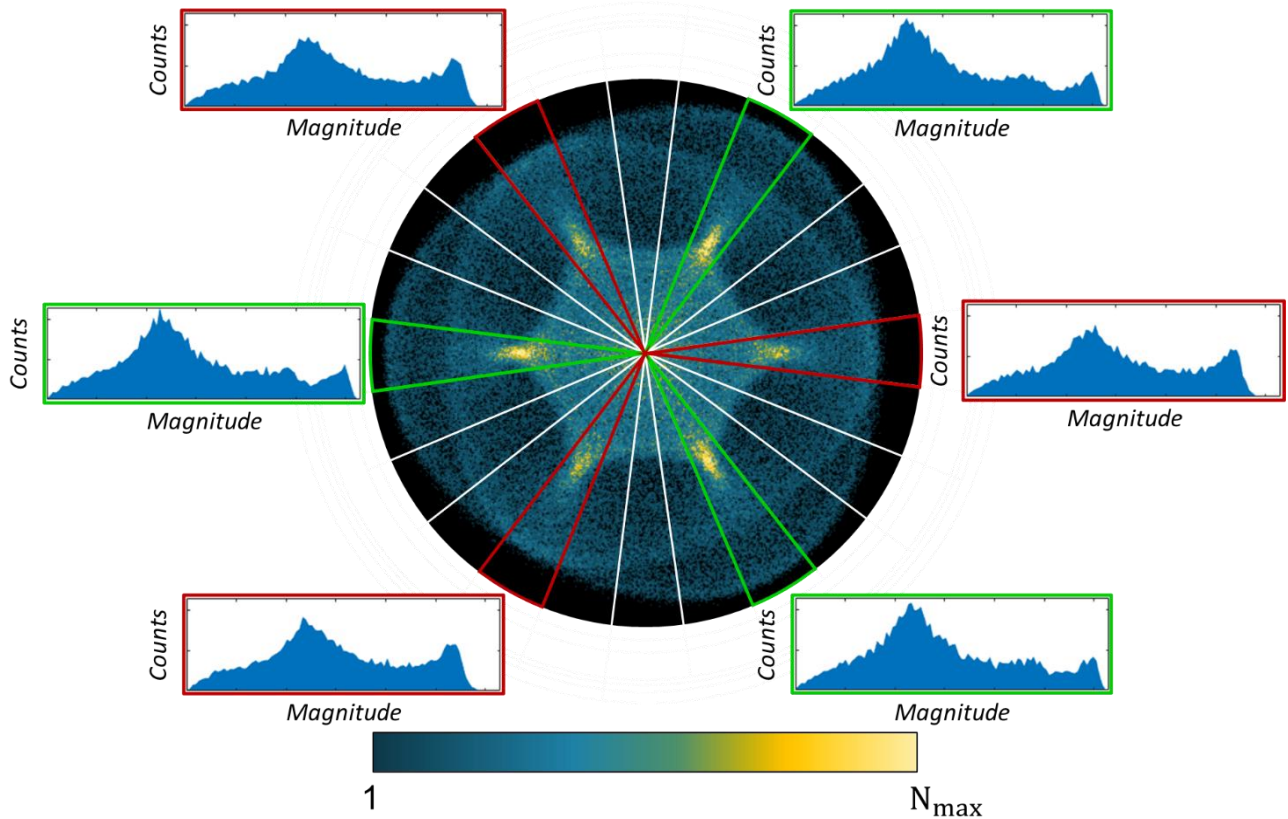

**Figure 3.** MoS$_2$ scattergram sectioned to show the slice configuration used in its analysis. The intensity profiles demonstrate how these can be used to validate the axes of symmetry from the Mo and S sub-lattices.

The influence of the instrumental parameters is quantitatively analyzed using the above decomposition scheme and comparing the maximum electrostatic field magnitudes. The quantification is performed by comparing the images under analysis to the aberration-free standard, determining the relative change induced in the maximum magnitude along every possible direction. Lower magnitude features are more susceptible to noise and therefore often not clearly identifiable in experimental scattergrams, hence, these are instead explored by qualitative observations.

*2.5.2. Electrostatic potential error maps*

The scalar electrostatic potential images are directly compared through relative difference maps that show the variation in magnitude at every pixel of the imaged region. Analysis of the influence of a parameter is based on the spatial distribution of induced changes, the maximum and minimum relative differences found, and the average relative difference over the entire image.

# 3. Results and Discussion

## 3.1. Convergence semi-angle and defocus

In STEM imaging, the selection of convergence semi-angle α is based on a compromise between spatial resolution and depth-of-field. As the probe is made more convergent, it becomes finer, and the area sampled at each scanning point becomes smaller. At the same time, the increased convergence reduces the dimensions of the beam crossover, after which it begins to diverge. Although a large depth-of-field is not a requirement for observing 2D materials, a small value increases the susceptibility of the image as it becomes more difficult to keep the entire material in focus. For this reason, the spatial resolution and depth of field are analyzed in combination.

We begin with a visual analysis and comparison of the electrostatic field configurations obtained with eDPC and eCoM. The 10-nm-spanning (from -5 nm to +5 nm) focal series of pristine $MoS_2$ for the two values of α considered in this study is shown in Figure 4, while the corresponding high-angle annular dark-field (HAADF) images can be seen in Figure S2. Generally, the in-focus ($C_1 = 0$) images clearly show the expected loss in spatial resolution with decreasing α, which is seen as a deformation of the atomic core shapes due to the blurring of the interatomic spacings. This effect is less pronounced in the eCoM images, where the atomic sites display a more uniform three-fold symmetry and are better defined relative to the corresponding eDPC images. For the cases where $C_1 \neq 0$, a strong dependence with α is observed. Considering α = 30 mrad, where the highest resolution is achieved, visual interpretation is severely affected already at $|C_1| = 3$ nm. The simulated images using 21 mrad show more robustness, still maintaining adequate interpretability at $|C_1| = 4$ nm. Additionally, overfocus is clearly more detrimental to the image quality, leading to quantitative changes that are evident in the measured electrostatic configurations. On the other hand, the difference in contrast between the Mo atoms and S columns becomes less discernible in underfocus conditions, indicating a greater impact to the distinction between atomic sites. The differences in uniformity of the atomic sites between eDPC and eCoM images identified above become more pronounced with the addition of defocus.

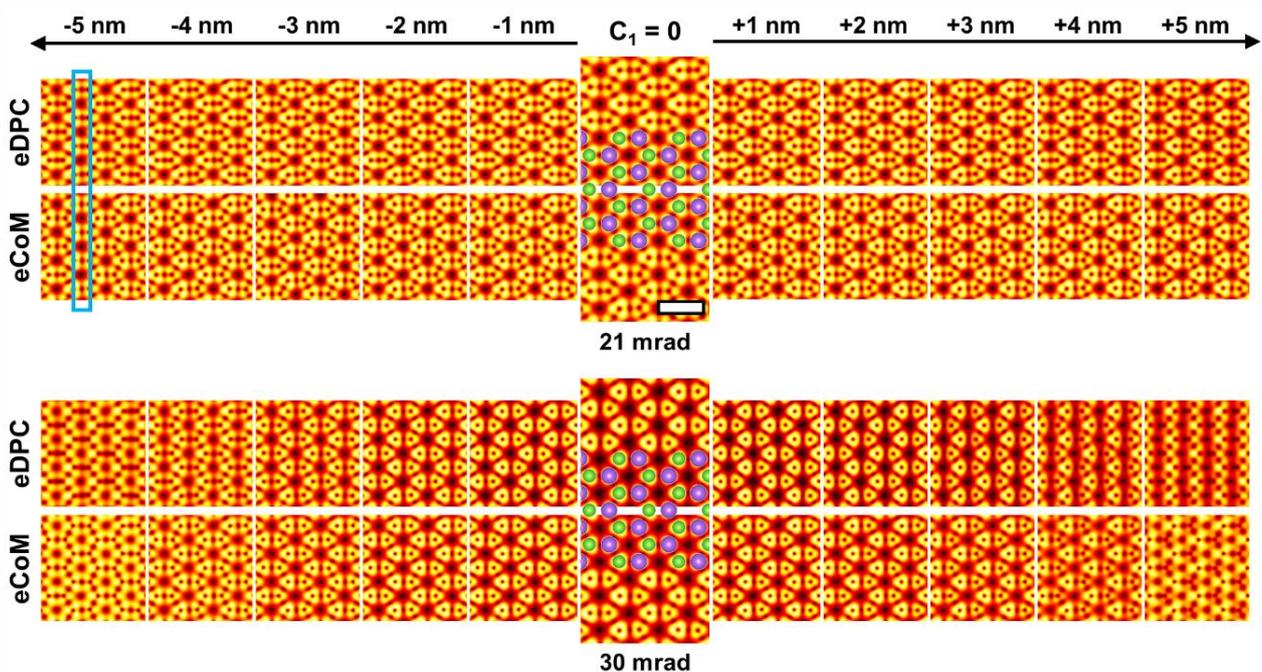

**Figure 4.** Collection of eDPC and eCoM images of pristine $MoS_2$ simulated using two convergence semi-angles (21 and 30 mrad) and a magnitude of defocus up to 5 nm at increments of 1 nm. Scale bar corresponds to 4 Å.

The impact of $C_1$ within each individual focus series is explored in detail through the relative difference graphs presented in Figure 5, which quantify the percentual change in maximum electrostatic field at every orientation (i.e. scattergram perimeter) compared to the corresponding aberration-free standard. The influence from the addition of $C_1$ on the eDPC images is made apparent in the corresponding graphs of Figure 5, which reach values beyond 60 % for 30 mrad, more than double the maximum values seen for 21 mrad. Note that this comparison is not equivalent to an analysis of absolute errors since, as mentioned above, the accuracy of CoM measurements is fundamentally limited by the probe convergence. In this case, the increased depth-of-field at 21 mrad adds flexibility to setting a proper focus, resulting in smaller percentual changes when defocus is introduced, as shown in the average percentual difference values plotted in Figure 6. A general trend is seen for both convergence angles, where the higher errors are observed along the directions of the segment edges, indicating that the CoM measurement close to these points is more affected by $C_1$. This effect is more noticeable for 30 mrad (due to the higher spatial resolution) and in the underfocused condition ($C_1 < 0$), which induces more qualitative changes. On the other hand, overfocused images are more qualitatively robust, showing a more rotationally symmetric error distribution, while the overall effect on the quantification is enhanced.

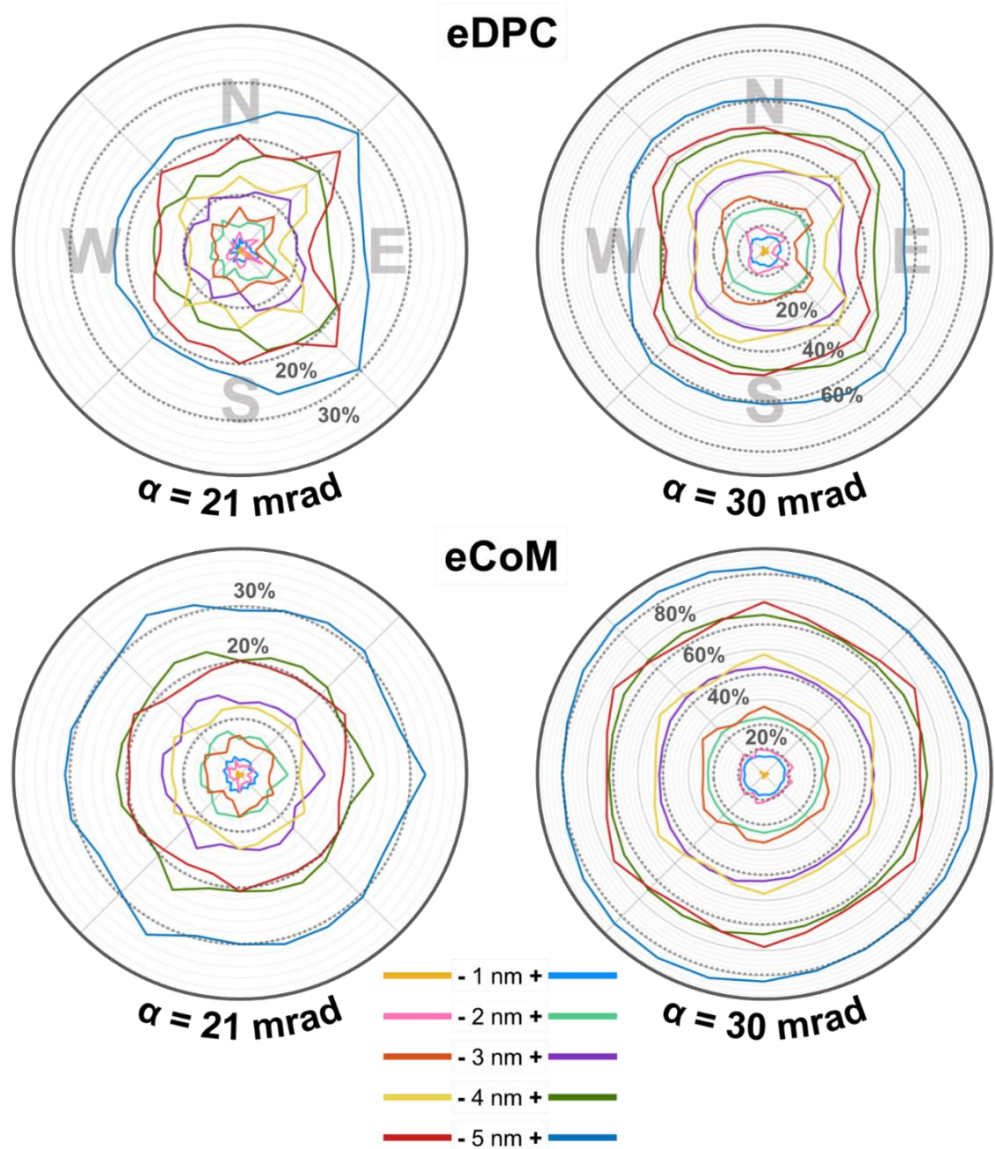

**Figure 5.** Relative eDPC and eCoM difference graphs comparing the distribution of maximum electrostatic field magnitude in images with added $C_1$ to that of the aberration-free standard.

For eCoM images, the relative difference between maximum percentual changes reached for both values of α is nearly identical to the segmented-detector images. However, a dependence on vector orientation is not seen, confirming that this is an effect specifically tied to a segmented detector geometry. The percentual changes across the $C_1$ series are therefore radially uniform for both α, showing that the aberration has a mostly quantitative influence on the high magnitude regions of the electrostatic configurations. Unlike in the segmented detector case, the more accurate detector type does not exacerbate the qualitive changes induced in underfocus but the quantitative impact associated with the overfocused condition is even clearer in the eCoM images, making this the most unfavorable situation when using a pixelated detector.

From the average relative differences of each aberrated image, shown in Figure 6, it is clear that operating in overfocus has a more detrimental effect to the quantification of both eDPC and eCoM measurements, causing changes over 10 % greater compared to those induced by underfocus at the interpretation limits. For negative $C_1$, the impact on the two techniques in the 21 mrad case is identical and only diverges slightly on the positive side, with eCoM showing higher changes in average. A larger divergence in seen for the 30 mrad probes, where eCoM shows higher susceptibility to the aberration. This discrepancy can be linked to a combination of the finer probe with the augmented sensitivity provided by the pixelated detector measurements.

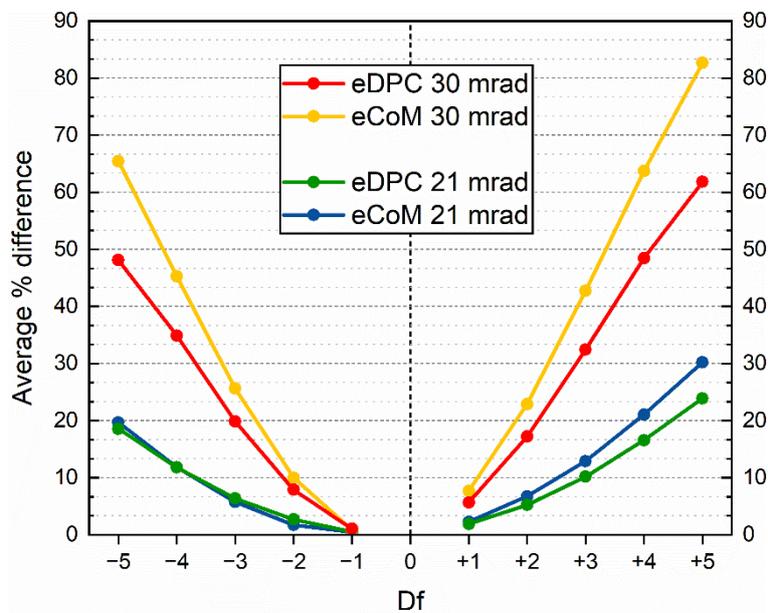

**Figure 6.** Average percentual difference in maximum electrostatic field magnitude for all directions relative to the value of $C_1$ for all combinations of detectors and semi-convergence angles considered.

Considering the results shown above for α = 21 mrad, an error around 16 % can be expected for the measurement of electrostatic field magnitudes at the limit of interpretability established at 4 nm of overfocus, while underfocus has a slightly decreased quantitative influence just above 10 %. In the case of α = 30 mrad, the quantitative changes from overfocus can easily surpass 30 % at the interpretability limit of 3 nm, whereas the influence of underfocus is around 20 %.

The impact of defocus $C_1$ on the electrostatic potential images is analyzed through the relative difference maps shown in Figure 7, where the color scales are limited to the modulus of the highest percentual difference found, on both the positive and negative sides.

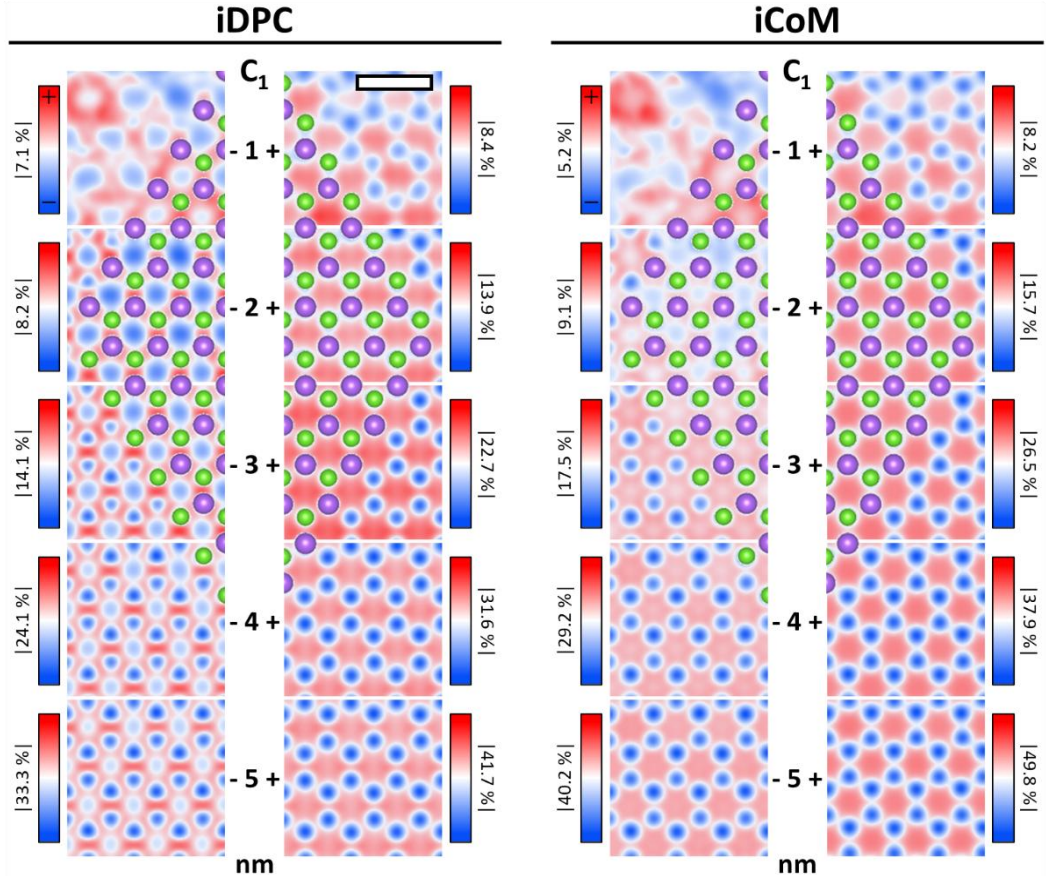

**Figure 7.** Relative difference iDPC and iCoM maps comparing the spatial configuration of electrostatic potential with added $C_1$ to those of the corresponding aberration-free standards. The color scales are bound by the highest absolute percentual difference found in each map, for ease in comparing the differences between over- and underestimated regions. Scale bar corresponds to 5 Å.

The segmented detector simulations again show that a larger convergence angle increases the susceptibility of the technique to uncorrected defocus, with maximum relative difference values being notably higher for 30 mrad. For a 1 nm magnitude of $C_1$, the difference maps show a non-periodic distribution of changes that is particularly evident for the underfocused probe, suggesting that the influence of $C_1$ at such magnitudes interacts with the influence of thermal vibrations introduced in the simulations. Consistently between different α and across the range of $C_1$, the atomic site regions suffer a decrease in potential relative to the corresponding aberration-free standards, with the S sites being the most affected. This indicates that uncorrected defocus generally causes an underestimation of positive potentials proportional to the magnitude of the real object. In underfocus, this influence is well localized to the atomic site regions, however, on the side of overfocus, the reach of the overestimated region is extended. For the lower value of α, this effect is more severe and results in a network of underestimated potential between the neighboring atomic sites. While only the positive atomic site regions are underestimated in overfocus, the lowest potential regions, i.e. the center of the hexagonal motifs, are also underestimated when in underfocus, an effect which becomes subdued in range and intensity as the magnitude of $C_1$ increases. For the 30 mrad probe, however, the impact in these regions is more intense, causing significant underestimation of nearly the entire area of the $MoS_2$ hexagonal ring centers at low magnitudes of $C_1$. In the overestimated regions, the influence of underfocus is noticeably less uniform, presenting several variations in intensity up to values close to those of the highest underestimations in the same image. The quantitative changes can be assessed by comparing the highest relative differences in each map, which shows that overfocus can lead to larger inaccuracies regarding the magnitude of the electrostatic potential, especially when using a larger α. As such, these observations match those from the electrostatic field analysis regarding the stronger influence of overfocus on the quantification.

The relative difference maps for iCoM measurements show similar trends to those identified from the iDPC images albeit with noticeably less influence on the accuracy of the measurements. Specifically, the underestimation at the center of the hexagonal rings previously seen in the underfocused images is greatly reduced, being almost non-existent for the lower α, making the error in these regions more uniform. The quantitative influence in the overestimated regions when in overfocus is also subdued, relative to the highest overestimation in the same image, this being particularly beneficial in the 21 mrad case. In general, the influence of $C_1$ in the iCoM images is more localized, with underestimations being limited to the atomic sites, reducing much of the divide between qualitative changes induced by under- and overfocus, as well as semi-convergence angles. However, the discrepancy between the influence on the Mo and $S_2$ atomic sites at underfocus is still significant, so that overfocus remains less impactful to the electrostatic potential configurations.

The quantitative changes induced by $C_1$ on the electrostatic potential images can be evaluated from Figure 8, where the highest relative difference values determined in each map of Figure 7 are plotted. The susceptibility of the larger α is promptly observed, starting at an error around 7 % larger than 21 mrad, that grows faster with higher $C_1$ magnitude. The stronger impact of overfocus in the quantification capability of the techniques is also readily observed by the steeper slopes on the positive side of $C_1$. The higher sensitivity of iCoM can still be observed, however it is significantly decreased in relation to the eCoM images, only becoming apparent at higher amounts of overfocus for the lower α. In general, the maximum difference values observed are also lower than the corresponding average differences found for the eDPC and eCoM images, indicating that the integration of field images reduces the quantitative changes induced by $C_1$.

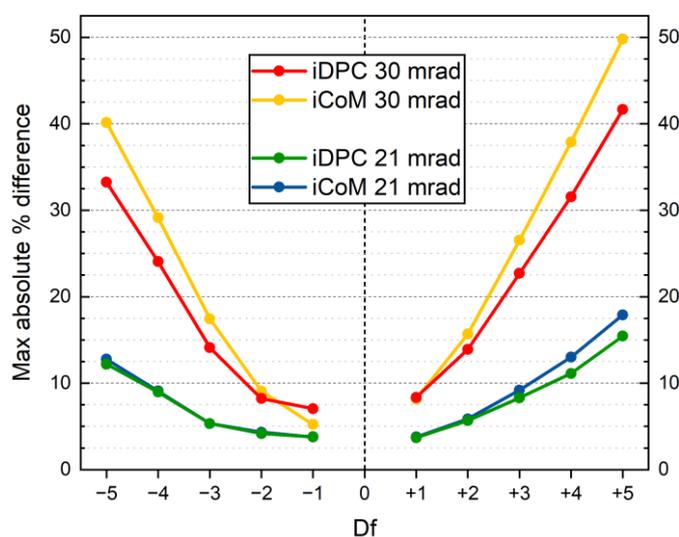

**Figure 8.** Maximum absolute percentual difference in electrostatic potential relative to the value of $C_1$ for all combinations of detectors and semi-convergence angles considered.

Based on the quantitative analysis, the influence of $C_1$ on electrostatic potential images can lead to overestimations of around 10 % for an α of 21 mrad at the 4 nm interpretability limit. For an α of 30 mrad, the quantitative changes from overfocus are above 20 % at the interpretability limit of 3 nm, with the influence of underfocus standing at a lower value below 15 %.

*3.2. Two-fold astigmatism*

Unlike the rotationally symmetric influence of $C_1$, the presence of $A_1$ induces distinct focal points between its two perpendicular axes, causing the image to appear stretched. In the following section, the orientation of this aberration is defined according to the underfocus axis of the aberration, following the system shown in Figure S3.

The influence of $A_1$ on electrostatic field maps is shown in Figure 9, where three different orientations for the aberration are considered up to a magnitude ($|A_1|$) of 3 nm - the point at which the MoS$_2$ atomic structure becomes barely identifiable. The corresponding HAADF images are shown in Figure S4. At an orientation $\theta_{A_1} = 0°$, the addition of astigmatism causes a blurring of the atomic site regions in the vertical direction, while low-field regions at the hexagonal ring centers become compressed in the perpendicular direction due to blurring of the atomic sites. At positions where the interatomic positions are vertically aligned, the atomic sites begin merging with the low-field divide, eventually splitting horizontally and creating a low-magnitude gap at the highest magnitude of $A_1$. While not visible in the segmented detector images, the eCoM configurations show that the low-field regions are mainly compressed in the horizontal direction, due to the vertical blurring of the surrounding high-magnitude atomic sites. With $A_1$ oriented at 60°, the axis of the aberration connects Mo and S$_2$ positions at opposite corners of the hexagonal ring (as identified by the overlaid arrow on the eDPC image for $|A_1| = 1$ nm in Figure 9). The influence of $A_1$ blurs the images in this direction, spreading the atomic site regions and causing a compression of the low-magnitude regions in the same direction due to the extended high-magnitude fields. While for $\theta_{A_1} = 0°$ this compression effect blurs together the magnitude from atomic sites along paths separated by the vertically aligned interatomic regions (i.e. three atoms for each hexagon), the alignment of $A_1$ with the diagonal of the hexagonal motif leads to the blurring of adjacent atomic site in pairs. This difference results in less qualitatively distorted atomic sites for $\theta_{A_1} = 60°$, progressively making their profiles seem more elongated along the $A_1$ axis with increasing magnitude of the aberration. The eCoM images for the 60° orientation are fairly similar, however the difference in magnitude between the Mo and S$_2$ atomic sites is increased. This results in the former becoming less visible as $A_1$ increases in intensity and the image becoming more blurred. The remaining orientation at 180° exhibits the opposite situation to that seen for 0°, with the atomic sites becoming horizontally smudged. This configuration of $A_1$ has a larger impact on image interpretability, although similar to the influence observed for the previous orientation at lower aberration magnitudes. At the highest magnitude considered, however, the low-field regions are more constricted and identification of atomic sites in eDPC images becomes nearly impossible as these become split in the horizontal direction. In the case of eCoM images, the low-field regions are also seen to go through a stronger constriction, but image interpretability is lost only up to the same degree as for the 60° orientation.

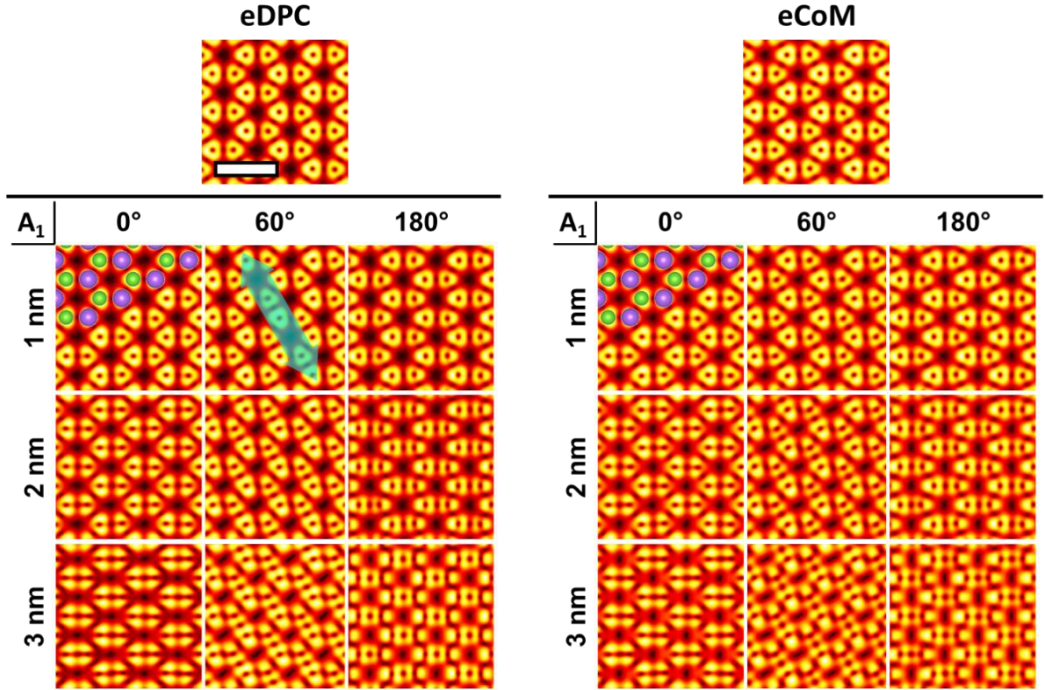

**Figure 9.** Collection of eDPC and eCoM images of $MoS_2$ simulated with the addition of $A_1$ in three orientations (0º, 60º and 180º) up to a magnitude of 3 nm at increments of 1 nm. Scale bar corresponds to 5 Å.

The relative percentual difference graphs analyzing the changes induced by $A_1$ on the maximum electrostatic field magnitude measured at every vector direction are presented in Figure 10. The graph corresponding to the eDPC scattergrams clearly shows the directional influence of the aberration, as relative differences increase along a specific axis for each orientation of $A_1$. The quantitative changes are also stronger than those induced by $C_1$, with maximum differences reaching close to 20 % for the lowest $A_1$ magnitude considered and values just below 80 % for the highest. Following the discussion above, the strongest influence is observed in the direction perpendicular to the $A_1$ axis where the field magnitude is most suppressed, while the error distribution in the remaining directions varies between orientation of $A_1$. The fact that similar variations are seen in the graph for the eCoM images indicates that these are connected to the structural characteristics of $MoS_2$, so that the influence of this aberration will differ depending on its orientation relative to the sample being observed. However, the variation in quantitative changes from different orientations in the eCoM images is minor compared to the eDPC case. While the strongest influence comes from $\theta_{A_1} = 0°$ in both detector types, the other two orientations show similar distributions that are around 10 % lower at the highest magnitude of $A_1$. In the eDPC simulations, $\theta_{A_1} = 60°$ always presents the lowest impact, while similar distributions are observed for the $\theta_{A_1} = 0°$ and $\theta_{A_1} = 180°$ orientations up to a $A_1$ magnitude of 2 nm. At a magnitude of 3 nm, only the 180º orientation results match between detector types, with changes induced at $\theta_{A_1} = 0°$ reaching values higher by more than 10 % while those at $\theta_{A_1} = 60°$ being inferior by around the same amount. These observations suggest an additional interplay between the orientation of $A_1$ and that of the segmented detector, which is explored further below.

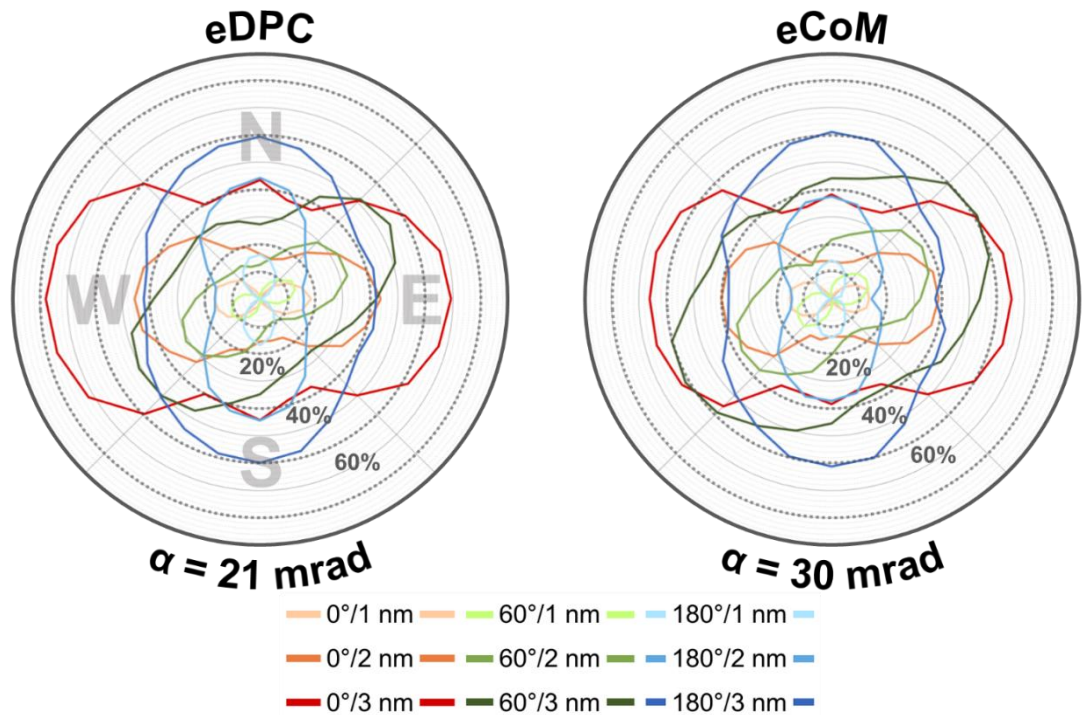

**Figure 10.** Relative eDPC and eCoM difference graphs comparing the distribution of maximum electrostatic field magnitude in images with added $A_1$ to those of the corresponding aberration-free standards.

The quantitative changes induced by $A_1$ can be compared in Figure 11. For a magnitude of 1 nm, the average quantitative change from $A_1$ for the configurations analyzed is generally found between 5 and 9 %. These values then increase rapidly with the magnitude of the aberration, more than doubling with every increment. The trends observed in the graph match the previous observations, showing how the influence determined for $\theta_{A_1} = 60º$ is consistently the lowest in the eDPC simulations, while the other two orientations present similar values up to the second increment of $A_1$. In contrast, the eCoM curves for $\theta_{A_1} = 60º$ and $\theta_{A_1} = 180º$ are in near perfect agreement, with only $\theta_{A_1} = 0º$ rising slightly more at the highest $A_1$ magnitude. At this point, average percentual changes for the eCoM images stand at around 50 % for every orientation, while values below 40 % to more than 55 % are seen in the eDPC results depending on the configuration.

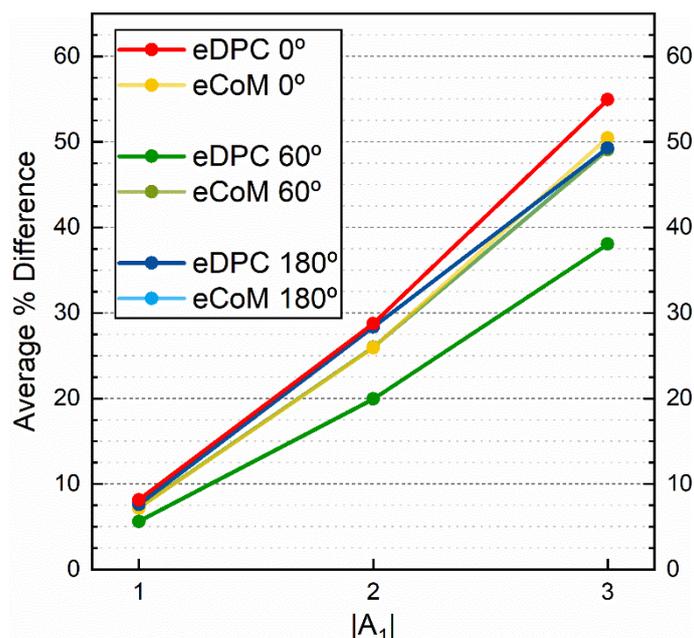

**Figure 11.** Average percentual difference in maximum electrostatic field magnitude relative to the magnitude of $A_1$ for all combinations of orientations and detectors considered.

Given the field-direction-biased influence of $A_1$, it is also important to have a reference for the maximum error induced by the aberration The graph in Figure S5 provides an indication of the quantitative changes that can be expected for the orientation at which the aberration is most detrimental, i.e. the direction perpendicular to its axis. The trends observed are similar to those of Figure 11, however the maximum percentual differences are substantially higher than the average values, even from the lowest aberration magnitude.

The influence of $A_1$ on electrostatic potential images is analyzed through the relative difference maps shown in Figure 12. The axial influence of the aberration is again seen in the iDPC difference maps, with underestimations concentrated to atomic sites and the interatomic regions connecting them in the direction of the $A_1$ axis. These interatomic regions show the smallest quantitative change, being those where the adjacent atomic potentials mutually cancel out, but that also create uneven distributions of quantitative changes that are harder to account for. In this sense, the difference maps for $\theta_{A_1} = 0º$ present the most predictable influence from the aberration, with under- and overestimations limited to atomic sites and interatomic spaces, respectively. The eCoM difference maps are comparable to those determined with the segmented detector configuration but show less overestimation in the interatomic regions and reduced quantitative changes around adjacent on-axis atomic sites.

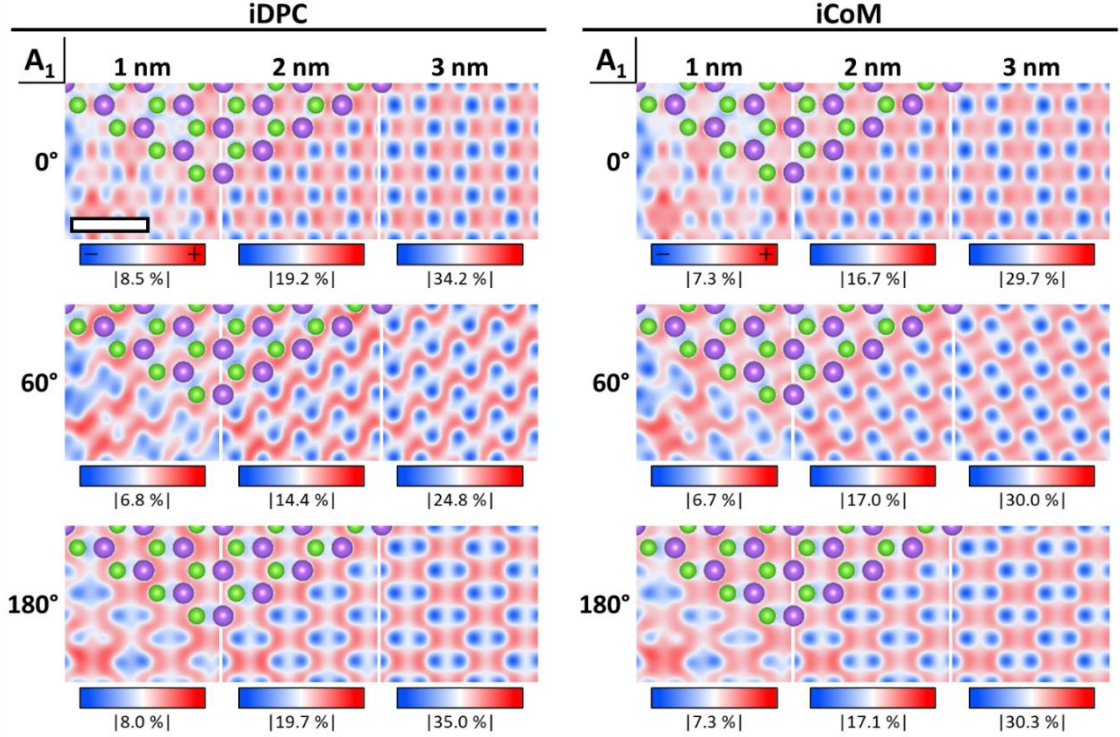

**Figure 12.** Relative difference iDPC and iCoM maps comparing the spatial configuration of electrostatic potential with added $A_1$ to those of the corresponding aberration-free standards. The color scales are bound by the highest absolute percentual difference found in each map, for ease in comparing the differences between over- and underestimated regions. Scale bar corresponds to 5 Å.

The extent of the quantitative changes induced by the aberration are compared in Figure 13, which shows the same trends observed in Figure 11. Overall, the $\theta_{A_1} = 60°$ orientation measured by the segmented detector results in the smallest influence among all datasets, while the other two orientations exhibit the highest. On the other hand, the quantitative changes induced in the iCoM images are constant, with all three curves overlapping. At the first increment of $A_1$, maximum difference values are only between 6 and 8 %, going through a more gradual growth with increasing magnitude. At a magnitude of 3 nm, the largest difference observed is found at 35 %, close to half of the highest average difference seen in the corresponding field images, demonstrating how the integration of field images also suppresses the quantitative influence of $A_1$.

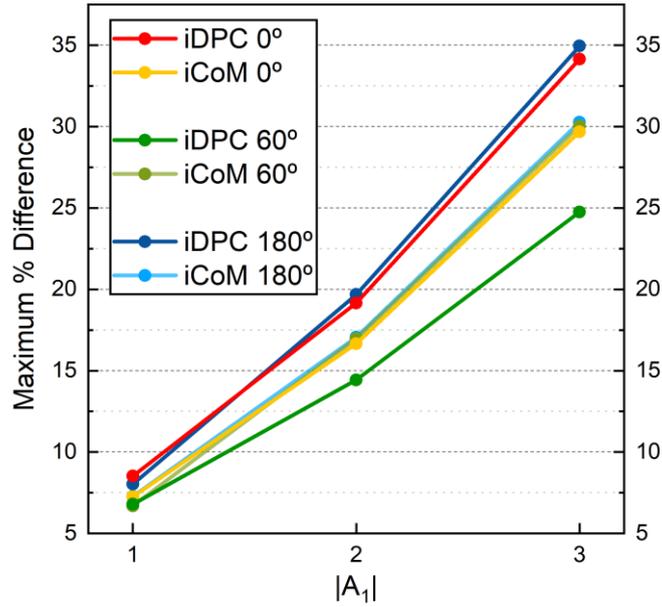

**Figure 13.** Maximum absolute percentual difference in maximum electrostatic potential relative to the magnitude of $A_1$ for all combinations of orientations and detectors considered.

A generalized expectation for the quantitative influence of $A_1$ on eDPC images is difficult to define given its dependence on the structure of the sample as well as the influence of the segmented detector orientation identified in the scattergram analysis. To explore this detector-related effect, the influence of $A_1$ is analyzed with respect to its orientation at a magnitude of 3 nm, the level at which the effect is observed most clearly (Figure 14). The graph in Figure 14A analyses the orientation dependence on eDPC images, showing a mostly parabolic distribution from vertical to horizontal alignment of the $A_1$ axis. The average differences decrease sharply towards the point at which the orientation of $A_1$ is aligned with the edges of the detector segments, dropping by about 15 % below the average eCoM value. The curve corresponding to the eCoM simulations provides a reference regarding the influence of sample structure for each $\theta_{A_1}$, showing that the average percentual changes are highest at orientations aligned with interatomic spacings (identified by $\Pi$ and $\Omega$ in Figure 14), following the previous observations where $\theta_{A_1} = 0°$ was identified as the most impactful orientation. Based on this behavior, the discrepant relative changes on the eDPC curve at $\theta_{A_1} = 0°$ and $\theta_{A_1} = 180°$ can be attributed to the influence of the sample structure. This indicates that the average error measured by a segmented detector will be highest when $A_1$ is oriented along the segments and lowest when oriented to the boundaries between them, where sensitivity is reduced, while the relative orientation of the structure modulates the overall effect on the images. The graph in Figure S6 shows the highest relative changes that can be expected for this aberration at every orientation. Analyzing this dependence on the maximum absolute difference in iDPC and iCoM images (Figure 14B) reveals a fairly constant curve for the iCoM simulations along the full range of orientations, such that the variations of the iDPC curve can be seen solely as a function of orientation of $A_1$ relative to the segmented detector. Here, a sinusoidal behavior is identifiable, with overestimated maxima at the segment midsections and underestimated minima at the edges. Based on the shape of this function as well as the fact that the underestimation at the minima is much larger, it can be concluded that the interaction between the orientation of $A_1$ and the segmented detector results from a lack of sensitivity in measuring the influence of the aberration when in near alignment with the segment edges. As $A_1$ is most detrimental along its two perpendicular axes, stronger changes become inaccurately measured as the aberration is oriented away from alignment with the segment midsections, leading to a decrease in the quantitative influence measured by the detector.

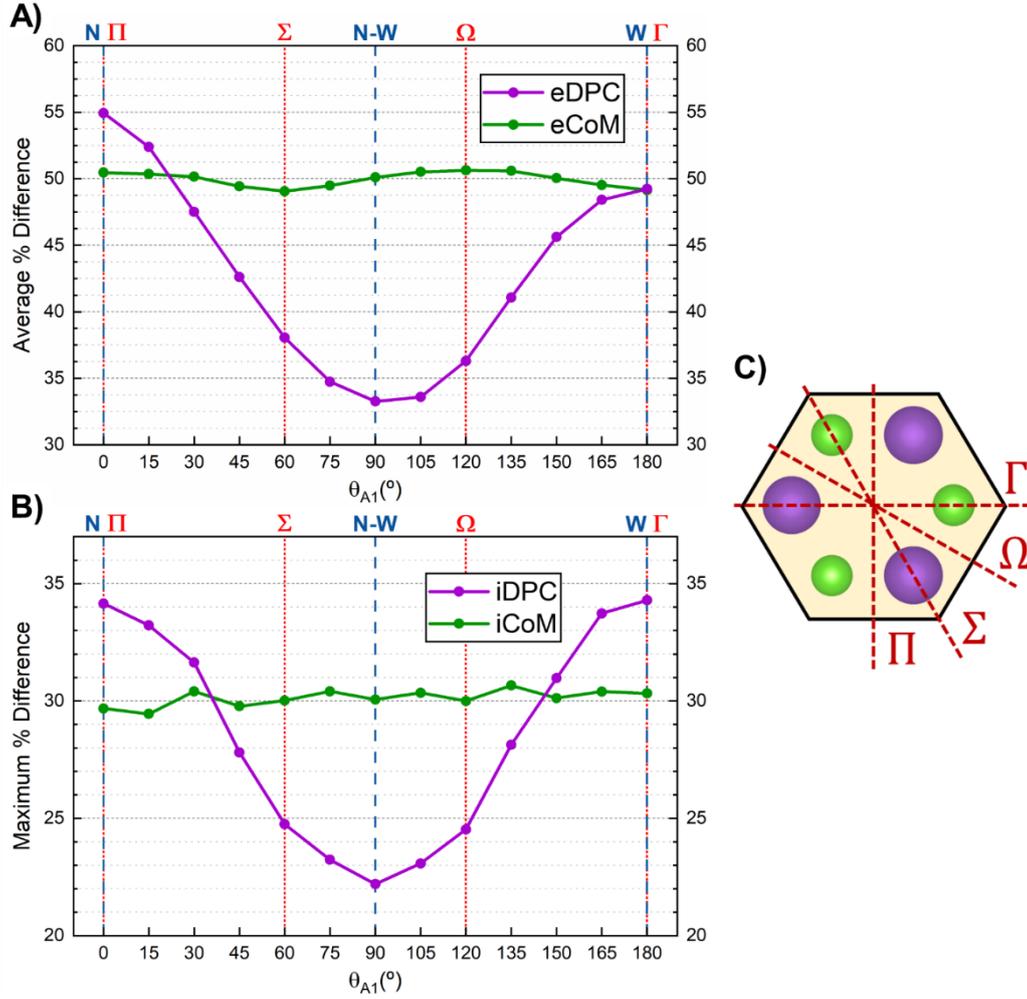

**Figure 14.** Error induced by $A_1$ on electrostatic mapping relative to its orientation for a magnitude of 3 nm and both types of detectors considered. A) Average percentual difference in maximum electrostatic field magnitude B) Maximum absolute percentual difference in electrostatic potential. C) Orientations identified in A) and B) relative to the $MoS_2$ structure.

## 4. Conclusions

The analysis performed in this study validates the usefulness and sensitivity of the scattergram-based approach to understand the overall qualitative and quantitative impact of electron optics aberrations in eDPC images. Using this methodology, the influence of semi-convergence angle and defocus was determined, finding that the underfocus condition – although less detrimental quantitatively – causes significative qualitative changes, leading to significant changes on the measured configurations. It was also found that the influence of $C_1$, though mostly radial, will be larger for field orientations closely aligned to the detector edges due to a lack of accuracy for measurements in such configurations. The impact of two-fold astigmatism analyzed under the same approach provided a perspective on the influence of non-rotationally symmetric aberrations, showing qualitative changes biased in the direction of its axes of symmetry. Additionally, the interaction between $A_1$ and the sample structure as well as the influence from the segmented detector geometry was identified by exploring the aberration along a range of orientation, which revealed a dip in sensitivity for orientations aligned with the edges of the detector segments. The impact of these instrumental parameters was also analyzed on iDPC images, demonstrating that the integration of eDPC images reduces their quantitative influence and suppresses some of the induced qualitative changes.

## Author contributions

R.V.F.: conceptualization, methodology, software, investigation, formal analysis, visualisation, writing - original draft, funding acquisition. S.C.V.: supervision, writing - reviewing and editing. P.J.F.: supervision, resources, writing – reviewing and editing, funding acquisition.

## Funding

R.V.F. acknowledges the funding from the Portuguese Science and Technology Foundation (FCT) through the PhD fellowship SFRH/BD/149390/2019. The authors also acknowledge FCT for its financial support via LAETA (project UID/50022/2025). This work was also supported by the AttoSwitch project (N°101135571) within the HORIZON-CL4-2023-DIGITAL-EMERGING-01-CNECT program, funded by the European Commission.

## Notes

The authors declare no competing interests.